\documentclass[aps,prd,preprintnumbers,amsmath,amssymb,showpacs,showkeys]{revtex4}

\usepackage[latin9]{inputenc}
\usepackage{color}
\usepackage{textcomp}
\usepackage{amsmath}
\usepackage{graphicx}
\usepackage{esint}
\usepackage{amssymb}
\usepackage{latexsym}
\usepackage{changes}
\usepackage{epsfig}

\usepackage{bm}
\usepackage{palatino}
\usepackage{physics}

\usepackage[colorlinks=true,linkcolor=blue,urlcolor=blue,citecolor=blue]{hyperref}

\begin{document}

\title{Schwarzschild-like black hole with a topological defect in bumblebee gravity}

\author{\.{I}brahim  G\"{u}ll\"{u}}
\email{ibrahim.gullu@emu.edu.tr}
\affiliation{Physics Department, Eastern Mediterranean
University, Famagusta, 99628 North Cyprus, via Mersin 10, Turkey}

\author{Ali \"{O}vg\"{u}n}
\email{ali.ovgun@emu.edu.tr}
\homepage[]{https://www.aovgun.com}
\affiliation{Physics Department, Eastern Mediterranean
University, Famagusta, 99628 North Cyprus, via Mersin 10, Turkey}

\date{\today}
\begin{abstract}
In this paper, we derive an exact black hole spacetime metric in the Einstein-Hilbert-Bumblebee (EHB) gravity around global monopole field. We study the horizon, temperature, and the photon sphere of the black hole. Using the null geodesics equation, we obtain the shadow cast by the Schwarzschild-like black hole with a topological defect in bumblebee gravity. Interestingly, the radius of shadow of the black hole increases with increase in the global monopole parameter. We also visualize the shadows and energy emission rates for different values of parameters. Moreover, using the Gauss-Bonnet theorem, we calculate the deflection angle in weak field limits and we discuss the possibility of testing the effect of global monopole field and bumblebee field, on a weak deflection angle. We find that the global monopole parameter and also Lorentz symmetry breaking parameter has an increasing effect on the deflection angle.

\end{abstract}


\maketitle
\section{Introduction}

Spontaneous symmetry breaking is a key notion in particle physics and the symmetry can be either a kind of internal symmetry or a symmetry associated with spacetime transformations. Breaking of internal symmetries spontaneously causes topological defects which are called global. A general property of spontaneously broken global symmetries is the Goldstone degree of freedom which  in the monopole case results in a divergence of the total global monopole energy. In order to remove this self-energy problem one way is to introduce the self gravity of global monopole which is also essential for astrophysical applications \cite{Bronnikov}.
Monopoles which are formed by gauge-symmetry breaking during the phase
transitions in the early universe can be the source of inflation \cite{Preskill,Barriola}.
Monopoles are one of the example of stable topological defects \cite{Vilenkin:1981zs,Vilenkin:1982hm}.
On the other hand, global monopoles result from a global symmetry
braking of global $O\left(3\right)$ symmetry into $U\left(1\right)$
in phase transitions in the universe \cite{Vilenkin:1981zs,Vilenkin:1982hm,Vilenkin:1984ib,Vilenkin:1981kz}.
The effect of these monopoles on Schwarzschild spacetime are discussed
in \cite{Dadhich} by analyzing the particle orbit and Hawking radiation.
On the other side, the Lorentz symmetry breaking (LSB) is associated
with the idea that the Quantum Gravity (QG) signals may emerge at
low energy scales \cite{Cas}. The naturelness and the possibility
of Lorentz symmetry breaking is discussed in the context of string
theory in \cite{Kosteleck=0000FD-1,Kosteleck=0000FD-2,Kosteleck=0000FD-3,Kosteleck=0000FD-4,Kostelecky:2000ku,Kostelecky:2018yfa,Kostelecky:2002ca,Lehnert:2004be,Bertolami:2003qs,Diaz:2013saa,Kostelecky:2000mm,Lehnert:2018lce,RalfLehnert:2016grl,Cambiaso:2012vb,Lehnert:2011ed,Bluhm:2004ep}.
The Lorentz symmety breaking arises in other theories like noncomutative
field theories \cite{Carroll,Mocioiu,Ferrari} and loop quantum gravity
theory \cite{Gambini,Ellis} among other scenarios \cite{Berger,Blas,Lehnert,Gorbunov,Maluf:2021lwh,Oliveira:2021abg,Jha:2020qkm,Delhom:2021lxs,Delhom:2020gfv,Nascimento:2014vva,Assuncao:2017tnz,Assuncao:2019azw,Jha:2020pvk,Khodadi:2021owg,Ding:2020kfr,Chen:2020qyp,Ali:2020ddi,Ding:2021iwv,Carvalho:2021jlp,Ding:2019mal,Jesus:2020lsv,Oliveira:2018oha}.

Einstein's general relativity (GR) describes gravitation at the classical level. On the other hand standard model (SM) of particle
physics brings all other interactions together below the quantum level.
At the Planck scale, $m_{p}=10^{19}\;GeV$, these theories expected
to merge into one single unified theory which may provide a successful
description of nature. Altough, confirmation of this idea is compelling
because experiments at the Planck scale are unfeasible, the effect
emerging from the underlying QG theory might be observable at the
low-energy scales \cite{Cimdiker:2021cpz}. One way to set the framework of Planck-scale signals
is violating relativity and these violations are related to the breaking
of Lorentz symmetry. Theories violate the Lorentz symmetry at the
Planck scale and contain both GR and SM are effective field theories
known as the SM-Extension (SME) by which the observable signals of
Lorentz violation can be described \cite{Kostelecky-1,Kostelecky-0}.

Bumblebee models are effective field theories which
has a vector field with a non-zero vacuum expectation value (VEV)
that sponaneously breaks Lorentz symmetry \cite{Kosteleck=0000FD-3,Kostelecky:2000mm,Kostelecky-0}.
These particular type of effective field theories named by Kosteleck\'y
as Bumblebee models inspired from the real-life bumblebees. Real-life
bumblebees have a large bodies but tiny wings which give rise to think
bumblebees can not fly for a long time, however they can successfully
\cite{Dickinson}. Similarly, Bumblebee models does not have local
$U\left(1\right)$ gauge symmetry and expected not to have propagating
(massless) vector modes. Nonetheless, in these models massless modes
arise as Nambu-Goldstone (NG) modes which propagates \cite{Bluhm}. In the present work we bring together spontaneous LSB scenario with spontaneous global symmetry violation in order to see the cosmological implications of global monopoles on Bumblebee models.

Einstein's theory of general relativity, which is a metric theory
of gravitation \cite{Einstein:1916vd}, has successfully passed through
many experimental tests. One of the important testing technique is
a gravitational lensing which helps us to understand galaxies, dark
matter, dark energy and the universe \cite{Abbott:2016blz,Akiyama:2019cqa,Bartelmann:1999yn}.
This technique firstly was used by Eddington \cite{Dyson:1920cwa},
afterwards, various works on gravitational lensing have been done
for black holes, wormholes, global monopoles and other objects \cite{Virbhadra:1999nm,Virbhadra:2002ju,Virbhadra:2007kw,Virbhadra:2008ws,Keeton:1997by,Bozza:2002zj,Bozza:2009yw,Chen:2009eu,Eiroa:2002mk}.
There are many method to calculate gravitational lensing \cite{Sharif:2015qfa,Cao:2018lrd,Bisnovatyi-Kogan:2017kii,BisnovatyiKogan:2010ar,Cunha:2018acu}.
Recently, new method is derived by Gibbons and Werner which the deflection
angle of light can be calculated from non-rotating asymptotically
flat spacetimes using the Gauss-Bonnet theorem on the optical geometry
of the black hole \cite{Gibbons:2008rj}, then it is extended to stationary
spacetimes by Werner \cite{Werner:2012rc}. Since then, many works
have been done, one can see \cite{Crisnejo:2019ril}-\cite{Li:2020dln}.

The aim of the manuscript is to obtain a Schwarzschild like solution
with global monopole of Einstein equation in the presence of spontaneous
Lorentz symmetry breaking. We then study the effect of global monopoles
and spontaneous Lorentz symmetry breaking on the shadow of black hole
and also deflection of light. We also discuss the effect of global
monopoles on Lorentz symmetry breaking.

The manuscript is organized as follows: In section II, we derive the
Einstein fields equations for Einstein-Hilbert-Bumblebee gravity around
a global monopole fields. In section III, we obtain new Schwarzschild-like
black hole solution for Einstein-Hilbert-Bumblebee (EHB) gravity around
a global monopole. In section IV, we calculate the shadow of the black
hole. Section V is devoted to computation of the weak deflection angle
by Schwarzschild-like black hole solution for EHB gravity around a
global monopole using the GBT. We conclude our results in section
VI.

\section{EHB Gravity Around a Global Monopole}

In this part we review the EHB gravity \cite{Cas} whose Lagrangian
density is introduced as follows

\begin{align}
\mathcal{L}_{B} & =\sqrt{-g}\left(\frac{1}{2\kappa}R+\frac{\xi}{2\kappa}B^{\mu}B^{\nu}R_{\mu\nu}-\frac{1}{4}B_{\mu\nu}B^{\mu\nu}-V\left(B^{\mu}B_{\mu}\pm b^{2}\right)\right)+\mathcal{L}_{M},\label{Lag_Bumblebee}
\end{align}
where $B^{\mu}$ is the bumblebee field with field strength tensor
$B_{\mu\nu}=\partial_{\mu}B_{\nu}-\partial_{\nu}B_{\mu}$, $V\left(B^{\mu}B_{\mu}\pm b^{2}\right)$
being the potential which has terms responsible for the spontaneous
Lorentz symmetry breaking, $b^{2}$ is a real positive
constant and $\xi$ is the coupling constant of nonminimal gravity-bumblebee
interaction. $\mathcal{L}_{M}$ is the Lagrangian density of matter,
which in our case is the global monopole, and others field contents
with their couplings to the bumblebee field. Although,
the action (\ref{Lag_Bumblebee}) does not seem to be free from the
Ostrogradsky instabilities because of the coupling $B^{\mu}B^{\nu}R_{\mu\nu}$,
and is forbidden by gauge invariance in usual Einstein-Maxwell electrodynamics,
it is consistent in LSB scenarios with the axial gauge condition $b_{\mu}A^{\mu}=0$
(here we set the gauge potential $A^{\mu}$ instead of $B^{\mu}$)
\cite{Bluhm:2004ep,Kostelecky-0} where $b^{\mu}$ is the vacuum expectation
value (VEV) of $B^{\mu}$, i.e. $\left\langle B^{\mu}\right\rangle =b^{\mu}$,
in the Minkowski background spacetime geometry. 
The equations of
motion of (\ref{Lag_Bumblebee}) is 
\begin{align}
\mathcal{G}_{\mu\nu}= & \kappa\left(T_{\mu\nu}^{B}+T_{\mu\nu}^{M}\right)\label{EoM_1}
\end{align}
after varying (\ref{Lag_Bumblebee}) with respect to the metric $g_{\mu\nu}$.
We defined the Einstein tensor $\mathcal{G}_{\mu\nu}=R_{\mu\nu}-\frac{1}{2}g_{\mu\nu}R$
and energy-momentum tensor $T_{\mu\nu}=T_{\mu\nu}^{B}+T_{\mu\nu}^{M}$
where $T_{\mu\nu}^{B}$ is the contribution of the bumblebee field
to the energy-momentum tensor 
\begin{eqnarray}
T_{\mu\nu}^{B}\equiv & -B_{\mu\sigma}B_{\phantom{\sigma}\nu}^{\sigma}-\frac{1}{4}g_{\mu\nu}B_{\alpha\beta}^{2}-g_{\mu\nu}V\left(B^{\mu}B_{\mu}\right)+4V^{\prime}B_{\mu}B_{\nu}+\frac{\xi}{\kappa}\left(\frac{1}{2}g_{\mu\nu}B^{\alpha}B^{\beta}R_{\alpha\beta}-B_{\nu}B^{\alpha}R_{\alpha\mu}-B_{\mu}B^{\alpha}R_{\alpha\nu}\right)\nonumber \\
 & +\frac{\xi}{\kappa}\left(\frac{1}{2}\nabla_{\alpha}\nabla_{\mu}\left(B^{\alpha}B_{\nu}\right)+\frac{1}{2}\nabla_{\alpha}\nabla_{\nu}\left(B_{\mu}B^{\alpha}\right)\right)+\frac{\xi}{\kappa}\left(-\frac{1}{2}\nabla^{\lambda}\nabla_{\lambda}\left(B_{\mu}B_{\nu}\right)-\frac{1}{2}g_{\mu\nu}\nabla_{\alpha}\nabla_{\beta}\left(B^{\alpha}B^{\beta}\right)\right),\label{T_mn^B}
\end{eqnarray}
and $T_{\mu\nu}^{M}$ is the matter sector of the energy-momentum
tensor which is defined as \cite{Barriola} 
\begin{align}
T_{\mu}^{\left(M\right)\nu}= & \text{diag}\left(\frac{\eta^{2}}{r^{2}},\frac{\eta^{2}}{r^{2}},0,0\right)\label{T_mn^M}
\end{align}
where $\eta$ is the constant term which is related to the global
monopole charge. Note that the total energy-momentum tensor is covariantly
conserved:
\begin{equation}
\nabla^{\mu}T_{\mu\nu}=  0.
\end{equation}
Taking the variation of (\ref{Lag_Bumblebee}) with respect to the
bumblebee field $B_{\mu}$ provides the second equation of motion
\begin{equation}
\nabla^{\mu}B_{\mu\nu}=J_{\nu}\label{EOM_2}
\end{equation}
where $J_{\nu}=J_{\nu}^{B}+J_{\nu}^{M}$, with $J_{\nu}^{M}$ acting
as a source term for the bumblebee field and $J_{\nu}^{B}=2V^{\prime}B_{\nu}-\frac{\xi}{\kappa}B^{\mu}R_{\mu\nu}$
is the current due to self interaction of the bumblebee field.

The trace of (\ref{EoM_1}) is 
\begin{align}
-\frac{1}{\kappa}R= & -4V\left(B^{\mu}B_{\mu}\right)+4V^{\prime}B_{\mu}B^{\mu}+\frac{\xi}{\kappa}\left(-\frac{1}{2}\nabla^{\lambda}\nabla_{\lambda}\left(B_{\mu}B^{\mu}\right)-\nabla_{\alpha}\nabla_{\beta}\left(B^{\alpha}B^{\beta}\right)\right)+T^{M}.\label{R}
\end{align}
Inserting the (\ref{R}) back into (\ref{EoM_1}) we get explicitly
\begin{align}
\frac{1}{\kappa}R_{\mu\nu}= & T_{\mu\nu}^{M}-\frac{1}{2}g_{\mu\nu}T^{M}-B_{\mu\sigma}B_{\phantom{\sigma}\nu}^{\sigma}-\frac{1}{4}g_{\mu\nu}B_{\alpha\beta}^{2}-g_{\mu\nu}V\left(B^{\mu}B_{\mu}\right)+4V^{\prime}B_{\mu}B_{\nu}+\frac{\xi}{\kappa}\left(\frac{1}{2}g_{\mu\nu}B^{\alpha}B^{\beta}R_{\alpha\beta}-B_{\nu}B^{\alpha}R_{\alpha\mu}-B_{\mu}B^{\alpha}R_{\alpha\nu}\right)\nonumber \\
 & +\frac{\xi}{\kappa}\left(\frac{1}{2}\nabla_{\alpha}\nabla_{\mu}\left(B^{\alpha}B_{\nu}\right)+\frac{1}{2}\nabla_{\alpha}\nabla_{\nu}\left(B_{\mu}B^{\alpha}\right)\right)+\frac{\xi}{\kappa}\left(-\frac{1}{2}\nabla^{\lambda}\nabla_{\lambda}\left(B_{\mu}B_{\nu}\right)-\frac{1}{2}g_{\mu\nu}\nabla_{\alpha}\nabla_{\beta}\left(B^{\alpha}B^{\beta}\right)\right)\nonumber \\
 & +\frac{1}{2}g_{\mu\nu}\left(4V\left(B^{\alpha}B_{\alpha}\right)-4V^{\prime}B^{\alpha}B_{\alpha}\right)-\frac{1}{2}g_{\mu\nu}\frac{\xi}{\kappa}\left(-\frac{1}{2}\nabla^{\lambda}\nabla_{\lambda}\left(B^{\alpha}B_{\alpha}\right)-\nabla_{\alpha}\nabla_{\beta}\left(B^{\alpha}B^{\beta}\right)\right)\label{R_mn}
\end{align}

The potential $V\left(B_{\mu}B^{\mu}\pm b^{2}\right)$ is chosen in
a way that the vacuum expectation value for $B_{\mu}$ has to be nonvanishing
in order to have spontaneous Lorentz symmetry breaking. Because of
this reason, the general form of the potential, $V\left(B_{\mu}B^{\mu}\pm b^{2}\right)$
is set to zero to determine the VEV of the bumblebee field which gives
the following condition 
\begin{align}
B_{\mu}B^{\mu}\pm b^{2}= & 0.\label{Condition}
\end{align}
The solution of (\ref{Condition}) has a nonnull vacuum expectation
value $\left\langle B^{\mu}\right\rangle =b^{\mu}$ which breaks the
Lorentz symmetry \cite{Cas}.

Note that Maxwell's electrodynamics in the non-linear gauge $A_\mu A_\mu-b^2 = 0$ and the bumblebee model with solutions confined to the case of $B_\mu B_\mu -b^2 = 0$ have same structure. However, in Maxwell's electrodynamics, the particles that carry that force, called photons, is understood by gauge symmetry, but in the bumblebee model is obtained from the nature of Goldstone bosons appearing from a spontaneous symmetry breaking of Lorentz invariance \cite{Escobar:2017fdi}.

\section{The Schwarzschild-like black hole of EHB Gravity Around Global Monopole}

In this part a static spherically symmetric solution to the Einstein
equations is obtained. We take the Birkhoff metric as an ansatz 
\begin{align}
g_{\mu\nu}= & \text{diag}\left(-e^{2\gamma},e^{2\rho},r^{2},r^{2}\sin^{2}\theta\right),\label{Bhirkoff}
\end{align}
where $\gamma$ and $\rho$ are functions of $r$, and fix the Bumblebee
field in its vacuum expectation value \cite{Bertolami:2005}

\begin{align}
B_{\mu}= & b_{\mu}.\label{B_equal_b}
\end{align}
Since the potential can be choosen simply as $V\left(x\right)=\frac{1}{2}\lambda x^{2}$
where $\lambda$ is a real coupling constant or $V\left(x\right)=\lambda x$
and as long as (\ref{Condition}) and (\ref{B_equal_b}) holds for
both choices of potentials one has \cite{Kostelecky-0}
\begin{align}
V= & 0,\quad V^{\prime}=0.\label{potential_0}
\end{align}
Also we take 
\begin{equation}
b_{\mu}=\left(0,b_{r}\left(r\right),0,0\right)\label{b_m}
\end{equation}
for which the field strength vanishes, $b_{\mu\nu}=0$. Although,
the ansatz (\ref{b_m}) can be generalized for $b_{t}\ne0$ (i.e.,
$b_{t}=\text{constant}$) for simplicity we keep $b_{t}=0$. Note that setting $B_t=0$ indicates that the vector field can diverge at the event horizon. One can also introduce $B_t=q=constant$, which still gives $B_{\mu \nu}=0$.  The advantage of such generalized solution is that the vector field is regular either the future or past event horizon \cite{Heisenberg:2017hwb}. Moreover, the case of $B^{\mu} B_{\mu}\neq0$ is also interesting, and need to be studied separately. These cases will be studied in our future projects. The condition $b^{\mu}b_{\mu}=b^{2}=\text{constant}$ gives the explicit
form of the radial background field  
\begin{align}
b_{r}\left(r\right)= & \left|b\right|e^{\rho}.\label{b_r}
\end{align}
Then, (\ref{R_mn}) can be written as following 
\begin{align}
0= & \bar{R}_{\mu\nu}=R_{\mu\nu}-\kappa\left(T_{\mu\nu}^{M}-\frac{1}{2}g_{\mu\nu}T^{M}\right)-\frac{\xi}{2}g_{\mu\nu}b^{\alpha}b^{\beta}R_{\alpha\beta}+\xi b_{\nu}b^{\alpha}R_{\alpha\mu}+\xi b_{\mu}b^{\alpha}R_{\alpha\nu}-\frac{\xi}{2}\nabla_{\alpha}\nabla_{\mu}\left(b^{\alpha}b_{\nu}\right)-\frac{\xi}{2}\nabla_{\alpha}\nabla_{\nu}\left(b_{\mu}b^{\alpha}\right)\nonumber \\
 & +\frac{\xi}{2}\nabla^{\lambda}\nabla_{\lambda}\left(b_{\mu}b_{\nu}\right).\label{Einstein_eqns}
\end{align}
where the trace of the energy-momentum tensor is \cite{Barriola}
\begin{align}
T^{M}= & 2\frac{\eta^{2}}{r^{2}}.\label{Trace_of_T_mn}
\end{align}
The combination constructed from the energy-momentum tensor and its
trace in (\ref{Einstein_eqns}) reads 
\begin{align}
T_{\mu\nu}^{M}-\frac{1}{2}g_{\mu\nu}T^{M}= & \left(0,0,-\eta^{2},-\eta^{2}\text{Sin}^{2}\left(\theta\right)\right),\label{T_mn_T}
\end{align}
and the components of the Ricci tensor in (\ref{Einstein_eqns}) are
\begin{align}
R_{tt}= & e^{2\left(\gamma-\rho\right)}\left[\partial_{r}^{2}\gamma+\left(\partial_{r}\gamma\right)^{2}-\partial_{r}\gamma\partial_{r}\rho+\frac{2}{r}\partial_{r}\gamma\right],\label{R_tt}\\
R_{rr}= & -\partial_{r}^{2}\gamma-\left(\partial_{r}\gamma\right)^{2}+\partial_{r}\gamma\partial_{r}\rho+\frac{2}{r}\partial_{r}\rho,\label{R_rr}\\
R_{\theta\theta}= & e^{-2\rho}\left[r\left(\partial_{r}\rho-\partial_{r}\gamma\right)-1\right]+1.\label{R_theta}
\end{align}
The components of (\ref{Einstein_eqns}) become 
\begin{align}
\bar{R}_{tt}= & \left(1+\frac{\ell}{2}\right)R_{tt}+\frac{\ell}{r}\left(\partial_{r}\rho+\partial_{r}\gamma\right)e^{2\left(\gamma-\rho\right)},\label{barR_tt}\\
\bar{R}_{rr}= & \left(1+\frac{3\ell}{2}\right)R_{rr},\label{barR_rr}\\
\bar{R}_{\theta\theta}= & \left(1+\ell\right)R_{\theta\theta}-\ell\left(\frac{1}{2}r^{2}e^{-2\rho}R_{rr}+1\right)+\eta^{2},\label{barR_theta}\\
\bar{R}_{\phi\phi}= & \sin^{2}\left(\theta\right)\bar{R}_{\theta\theta},\label{barR_phi}
\end{align}
where $\ell=\xi b^{2}$. These equations (\ref{barR_tt},\ref{barR_rr},\ref{barR_theta},\ref{barR_phi})
are indipendently equal to zero. Therefore, the following combination
can be written to find the function $\rho\left(r\right)$ \cite{Cas}
\begin{align}
r^{2}e^{-2\rho}\bar{R}_{tt}+r^{2}e^{-2\rho}\bar{R}_{rr}+2\bar{R}_{\theta\theta}= & 0,\label{first_comb}
\end{align}
which yields 
\begin{align}
e^{2\rho}= & \left(1+\ell\right)\left(1+\eta^{2}-\frac{\rho_{0}}{r}\right)^{-1}.\label{rho_func}
\end{align}
In order to find the function $\gamma\left(r\right)$ the following
second combination is constructed \cite{Cas} from (\ref{barR_tt},\ref{barR_rr},\ref{barR_theta},\ref{barR_phi})
which is 
\begin{align}
r^{2}e^{-2\gamma}\bar{R}_{tt}-\left(1+\frac{2}{\ell}\right)\bar{R}_{\theta\theta}= & 0.\label{second_comb}
\end{align}
After a straightforward calculation one reach the following function
\begin{align}
e^{2\gamma}= & 1+\eta^{2}-\frac{\rho_{0}}{r}.\label{gamma_func}
\end{align}
Then the Lorentz symmetry breaking spherically symmetric solution
for EHB can be written as 
\begin{align}
ds^{2}= & -\left(1-\bar{\mu}-\frac{2M}{r}\right)dt^{2}+\left(1+\ell\right)\left(1-\bar{\mu}-\frac{2M}{r}\right)^{-1}dr^{2}+r^{2}d\theta^{2}+r^{2}\sin^{2}\theta\,d\phi^{2}\label{solution}
\end{align}
where we have defined $\rho_{0}=2M$ and the global monopole term
$\bar{\mu}$ which is defined as $\bar{\mu}=-\eta^{2}$. The metric
(\ref{solution}) recovers LSB spherically symmetric solution when
$\eta=0$ \cite{Cas} and the usual Schwarzschild metric for limit
$\ell\rightarrow0$.

Then we introduce the following coordinate transformation: $r\rightarrow\left(1-\bar{\mu}\right)^{-1/2}r,t\rightarrow\left(1-\bar{\mu}\right)^{1/2}t,M\rightarrow\left(1-\bar{\mu}\right)^{-3/2}M$:
The spacetime metric of the Lorentz symmetry breaking spherically
symmetric solution for EHB becomes: 
\begin{align}
ds^{2}= & -\left(1-\frac{2M}{r}\right)dt^{2}+\left(1+\ell\right)\left(1-\frac{2M}{r}\right)^{-1}dr^{2}+g^{2}r^{2}\left(d\theta^{2}+\sin^{2}\theta\,d\phi^{2}\right)\label{solution}
\end{align}
where $g^{2}=1-\bar{\mu}$. 

The event horizon of the black hole locates at $g_{tt}(r_{h})=0$
so that $r_{h}=2M$, which does not depend on $\ell$ and $g$. It
means that there is not any effect on event horizon by bumblebee field
and global monopole field. On the other hand, Kretschmann invariant
is calculated as:

\begin{equation}
K_{\text{Kretschmann}}=R_{\alpha\beta\mu\nu}R^{\alpha\beta\mu\nu}={\frac{4\,\left({g}^{2}-\ell-1\right)^{2}{r}^{2}-16\,M{g}^{2}\left({g}^{2}-\ell-1\right)r+48\,{M}^{2}{g}^{4}}{{r}^{6}{g}^{4}\left(1+\ell\right)^{2}}},\label{KK}
\end{equation}
note that there is a physical singularity at $r=0$, because
the Kretschmann invariant is divergent.  
The Eq. \ref{KK} is different than the Schwarzschild case, which
means that the Eq. \ref{solution} is a real solution containing Lorentz-violating
corrections, because it is impossible to find a coordinate transformation
between the Eq. \ref{solution} and the Schwarzschild case. Moreover,
for $r_{h}=2M$ the Kretschmann scalar is finite therefore such a
singularity at the event horizon can be removed by means of a coordinate
transformation. The Hawking-Bekenstein temperature of the black hole
is \cite{Kanzi:2019gtu} 
\begin{equation}
T_{H}=\frac{\kappa}{2\pi}=\left.\frac{1}{4\pi\sqrt{-g_{tt}g_{rr}}}\frac{dg_{tt}}{dr}\right|_{r=r_{h}}=\left.\frac{1}{2\pi\sqrt{1+\ell}}\frac{M}{r^{2}}\right|_{r=r_{h}}=\frac{1}{8\pi M\sqrt{1+\ell}}
\end{equation}
where the surface gravity $\kappa=\nabla_{\mu}\chi^{\mu}\nabla_{\nu}\chi^{\nu}$
with Killing vector field $\chi^{\mu}$ and $k^{2}=-\frac{1}{4}g^{tt}g^{ab}g_{tt,a}g_{tt,b}$
is used. It is clear that the increasing LSB parameter $\ell$
reduces the temperature of the black hole and temperature does not
depend on the global monopole parameter $g$.



\section{Shadow of the Schwarzschild Like Solution with Global Monopole in
EHB Gravity}

Here, we investigate the shadow radius of the Schwarzschild like solution
with global monopole in EHB gravity. The Lagrangian $\mathcal{L}(x,\dot{x})=(1/2)g_{\mu\nu}\dot{x}^{\mu}\dot{x}^{\nu}$
for the geodesics of spherically symmetric and static spacetime metric
is 
\begin{equation}
\begin{aligned}\mathcal{L}(x,\dot{x})=\frac{1}{2}\left(-f(r)\dot{t}^{2}+\frac{(1+\ell)}{f(r)}\dot{r}^{2}+g^{2}r^{2}\left(\dot{\theta}^{2}+\sin^{2}\theta\dot{\phi}^{2}\right)\right)\end{aligned}
\end{equation}
where $f(r)=1-\frac{2M}{r}$.

Using the Euler-Lagrange equation $\frac{d}{d\lambda}\left(\frac{\partial\mathcal{L}}{\partial\dot{x}^{\mu}}\right)-\frac{\partial\mathcal{L}}{\partial x^{\mu}}=0$
within equatorial plane, ($\theta=\pi/2)$, the two conserved quantities
such as energy and angular momentum are obtained as

\begin{equation}
E=f(r)\dot{t},\quad L=g^{2}r^{2}\dot{\phi}.
\end{equation}

To obtain the geodesics equation for light, we write

\begin{equation}
0=-f(r)\dot{t}^{2}+\frac{(1+\ell)}{f(r)}\dot{r}^{2}+g^{2}r^{2}\dot{\phi}^{2}.
\end{equation}
Then using the conserved quantities $E$ and $L$ in to the above
equation, we find the orbit equation for photon as follows:

\begin{equation}
\left(\frac{dr}{d\phi}\right)^{2}=\frac{g^{2}r^{2}f(r)}{(1+\ell)}\left(\frac{g^{2}r^{2}}{f(r)}\frac{E^{2}}{L^{2}}-1\right),\label{eff}
\end{equation}
one can write it in the form of effective potential 
\begin{equation}
\left(\frac{dr}{d\phi}\right)^{2}=V_{eff}
\end{equation}

with 
\begin{equation}
V_{eff}=\frac{g^{4}r^{4}}{(1+\ell)}\left(\frac{E^{2}}{L^{2}}-\frac{f(r)}{g^{2}r^{2}}\right).
\end{equation}

The orbit equation depends only on the impact parameter $b=L/E$.
At the turning point of the trajectory $r=r_{ph}$, the condition
must satisfy $dr/\left.d\phi\right|_{r_{ph}}=0$ or $V_{eff}=0,\quad V_{eff}^{\prime}=0$
\cite{synge,Luminet:1979nyg}. The impact parameter at turning point
is

\begin{equation}
\frac{1}{b^{2}}=\frac{f(r_{ph})}{g^{2}r_{ph}^{2}}\label{impact}
\end{equation}

and using the conditions of $dr/\left.d\phi\right|_{r_{ph}}=0$ and
$d^{2}r/d\phi^{2}|_{r_{ph}}=0$, one can find the radius of the photon
sphere $r_{ph}$ by solving this equation: 
\begin{equation}
0=\frac{d}{dr}B(r)^{2}
\end{equation}
\begin{equation}
\frac{f^{\prime}\left(r_{ph}\right)}{f\left(r_{ph}\right)}-\frac{h^{\prime}\left(r_{ph}\right)}{h\left(r_{ph}\right)}=0.\label{photon}
\end{equation}
where $B^{2}(r)=\frac{h(r)}{f(r)}$ and $h(r)=g^{2}r^{2}$. Hence,
the analysis of Eq. \ref{impact} and Eq. \ref{photon} show that
the position of the photon sphere is $r_{ph}=3M$ and the critical
impact factor is $b_{crit}=3\sqrt{3}gM$.

In spherical symmetric spacetime, the black hole shadow is obtained
by using the light rings. The light rings corresponds to a critical
point of Eq. \ref{photon}. The Eq. \ref{eff} becomes 
\begin{equation}
\left(\frac{dr}{d\phi}\right)^{2}=\frac{h(r)f(r)}{1+\ell}\left(\frac{B^{2}(r)}{B^{2}(r_{ph})}-1\right).
\end{equation}
To calculate shadow radius, we use the angle $\alpha$ between the
light ray and radial direction as follows \cite{Perlick:2021aok}:
\begin{equation}
\cot\alpha=\left.\frac{\sqrt{(1+\ell)}}{\sqrt{f(r)h(r)}}\frac{dr}{d\phi}\right|_{r=r_{0}}.
\end{equation}
and 
\begin{equation}
\cot^{2}\alpha=\frac{B^{2}\left(r_{0}\right)}{B^{2}(r_{ph})}-1
\end{equation}
Using the relation $\sin^{2}\alpha=\frac{1}{1+\cot^{2}\alpha}$, we
obtain 
\begin{equation}
\sin^{2}\alpha=\frac{B^{2}(r_{ph})^ {}}{B^{2}\left(r_{0}\right)^ {}}.
\end{equation}
Since the critical value of impact parameter is related with radius
of photon sphere $b_{cr}=B(r_{ph})$, then the shadow is obtained
as

\begin{equation}
\sin^{2}\alpha_{\mathrm{sh}}=\frac{b_{\mathrm{cr}}^{2}}{B\left(r_{0}\right)^{2}}.
\end{equation}

Then the shadow radius of the black hole for a static observer $r_{0}$
is \cite{Konoplya:2019sns} 
\begin{equation}
R_{s}=r_{0}\sin\alpha=\sqrt{\frac{h(r_{ph})f\left(r_{0}\right)}{f\left(r_{ph}\right)}}
\end{equation}
and for a static observer at large distance is 
\begin{equation}
R_{s}^{2}=\frac{g^{2}r_{ph}^{2}}{f\left(r_{ph}\right)}.
\end{equation}
The apparent shape of the shadow can be found by a stereographic projection
in terms of celestial coordinates X and Y which as 
\begin{equation}
\begin{gathered}X=\lim_{r_{0}\longrightarrow\infty}\left(-\left.r_{0}^{2}\sin\theta_{0}\frac{d\phi}{dr}\right|_{\left(r_{0},\theta_{0}\right)}\right)\\
Y=\lim_{r_{0}\longrightarrow\infty}\left(\left.r_{0}^{2}\frac{d\theta}{dr}\right|_{\left(r_{0},\theta_{0}\right)}\right).
\end{gathered}
\end{equation}

\begin{figure}[ht!]
\centering \includegraphics[scale=0.6]{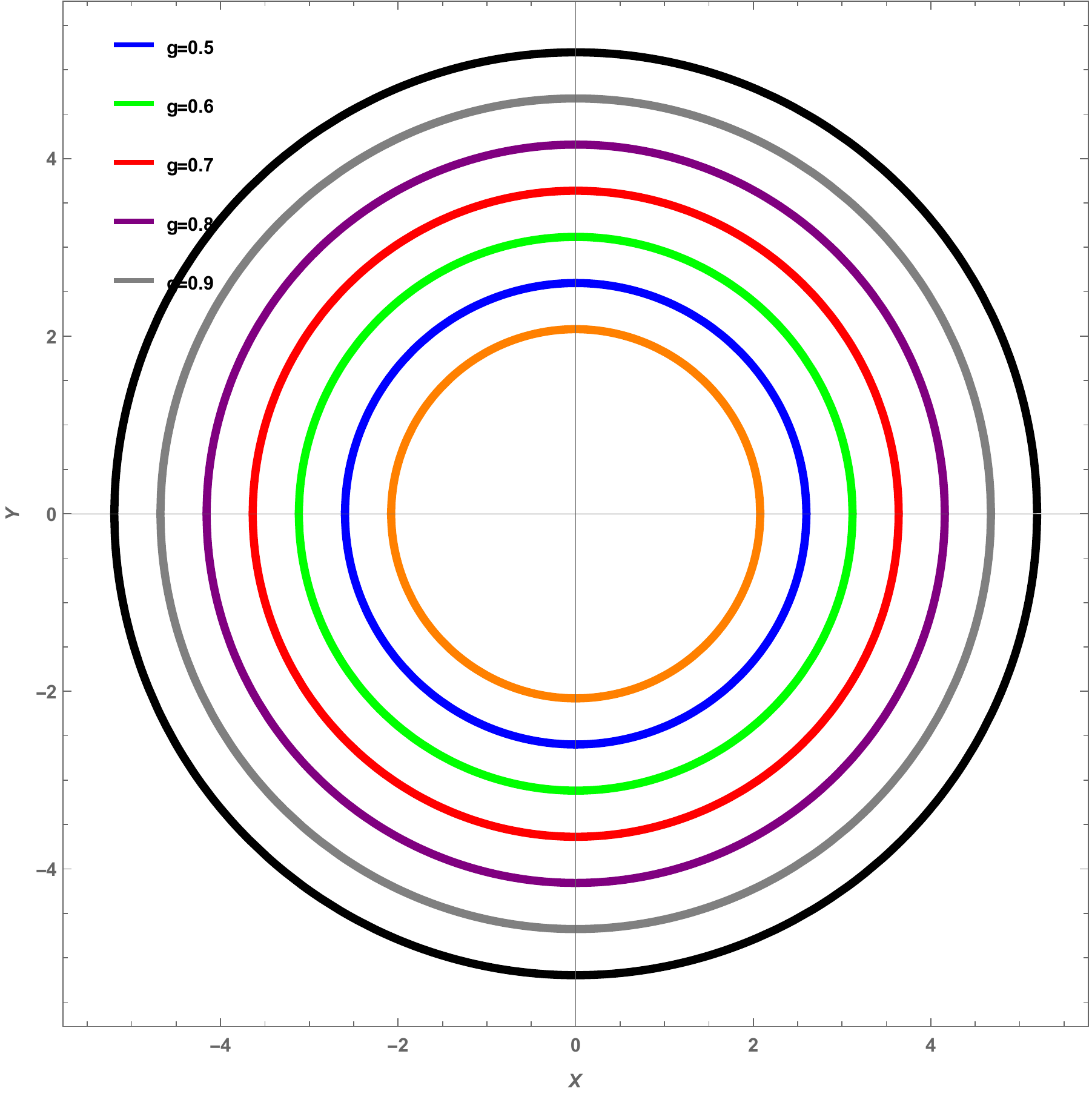} \caption{Shadow radius for Schwarzschild Like Solution with Global
Monopole in Bumblebee Gravity, $g=1$ Schwarzschild (black), $g=0.4$
(orange), $g=0.5$ (blue), $g=0.6$ (green), $g=0.7$ (red), $g=0.8$
(purple), and $g=0.9$ (gray) at $M=1$.}
\label{shadow} 
\end{figure}


Here, we shall use the  reported angular size of the black hole shadow in the M87$^*$ galactic center reported by  EHT  $\theta_s = (42 \pm 3)\mu as$, along with the distance to M87$^*$ given by  $D = 16.8 $ Mpc, and the mass of M87$^*$ central object $M = 6.5 \times 10^9$ M\textsubscript{\(\odot\)} to constrain the mass around the black hole M87$^*$. The diameter of the shadow in units of mass $d_{M87^*}$ is given by \cite{Allahyari:2019jqz}
\begin{eqnarray}
d_{M87^*}=\frac{D \,\theta_s}{M_{87^*}}=11.0 \pm 1.5,
\end{eqnarray}
where $d_{M87^*}=2 R_{M87^*}$. The observable quantity such as the angular diameter of the
M87$^*$  BH can be estimated using the observable $R_{M87^*}$
as follows $\theta_{s}=2 R_{M87^*} M_{M87^*} / D$.
The radius of the observed shadow of the M87$^*$ in unit mass is

\begin{eqnarray}
 R_{M87^*} = 5.5 \pm 0.75.
\end{eqnarray}
Shadow radius of the Schwarzschild-like black hole with a topological defect in bumblebee gravity  where the global monopole parameter is $\bar{\mu}\simeq10^{-5}$
\cite{Vilenkin:2000jqa} is 
\begin{eqnarray}
 R_{s} = 5.19613.
\end{eqnarray}
It is clear that Schwarzschild-like black hole with a topological defect in bumblebee gravity's shadow can appear up to $5.5\%$ smaller than it would in a  case of Schwarzschild-like black hole, providing a clear way to test topological defect.

In the Table.\ref{tab:table1}, we provide the shadow radius of the M87$^*$ as compared with Schwarzschild-like black hole with a topological defect in bumblebee gravity.
\begin{table}[ht!]
    \centering
    \begin{tabular}{ |p{2cm}|p{2cm}|p{2cm}|p{2cm}|p{2cm}|  }
    \hline
        $g$ & $R_s$ \\ [1 ex] 
        \hline
         $0.4 $ & $2.07846$ \\
        \hline
        $0.5 $ & $2.59808$ \\
        \hline
        $0.6$ & $3.11769$  \\
        \hline
        $0.7 $ & $3.63731$ \\
        \hline
        $1 $ & $5.19615$  \\
        \hline
    \end{tabular}
    \caption{Effect of the topological defects on the shadow size for the observed M87$^*$.}
    \label{tab:table1}
\end{table}

The shadow radius of the black hole increasing with the increasing
value of global monopole parameter $g$ as shown in Fig. \ref{shadow}.
On the other hand, the shadow radius does not depend on the LSB parameter
$\ell$. The relation between the high energy absorption cross section
and the shadow for the observer located at infinity is given by the
energy emission rate of the black hole as follows \cite{Decanini:2011xw,Wei:2013kza}:
\begin{equation}
\dfrac{d^{2}E(\omega)}{d\omega dt}=\dfrac{2\pi^{3}R_{s}^{2}}{e^{\omega/T}-1}\omega^{3},
\end{equation}

\begin{figure}[ht!]
\centering \includegraphics[scale=0.6]{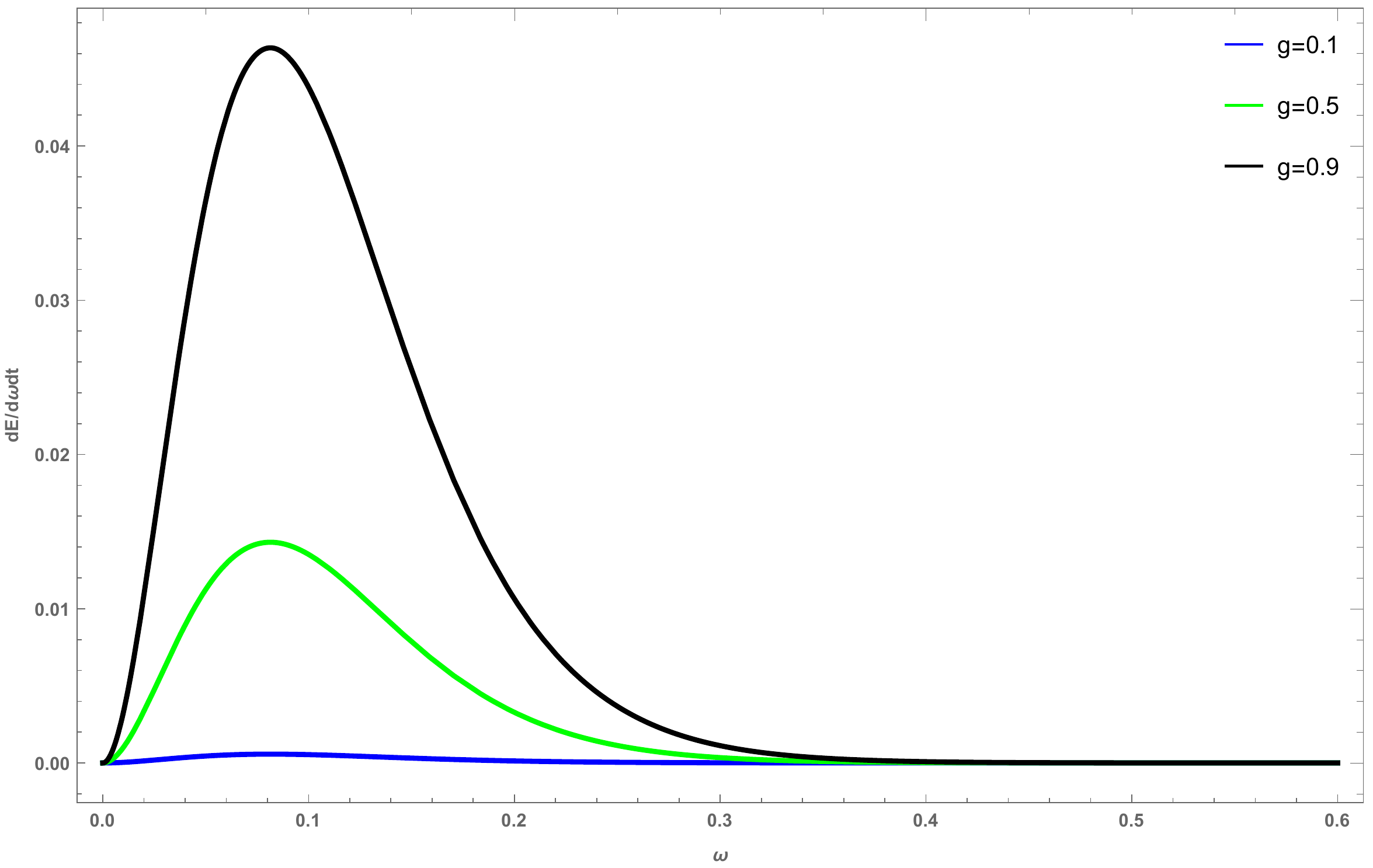} \caption{The BH emission rate for different values of magnetic monopole term
$g$ at $M=1$ and $\ell=0.9$.}
\label{emg} 
\end{figure}

\begin{figure}[ht!]
\centering \includegraphics[scale=0.6]{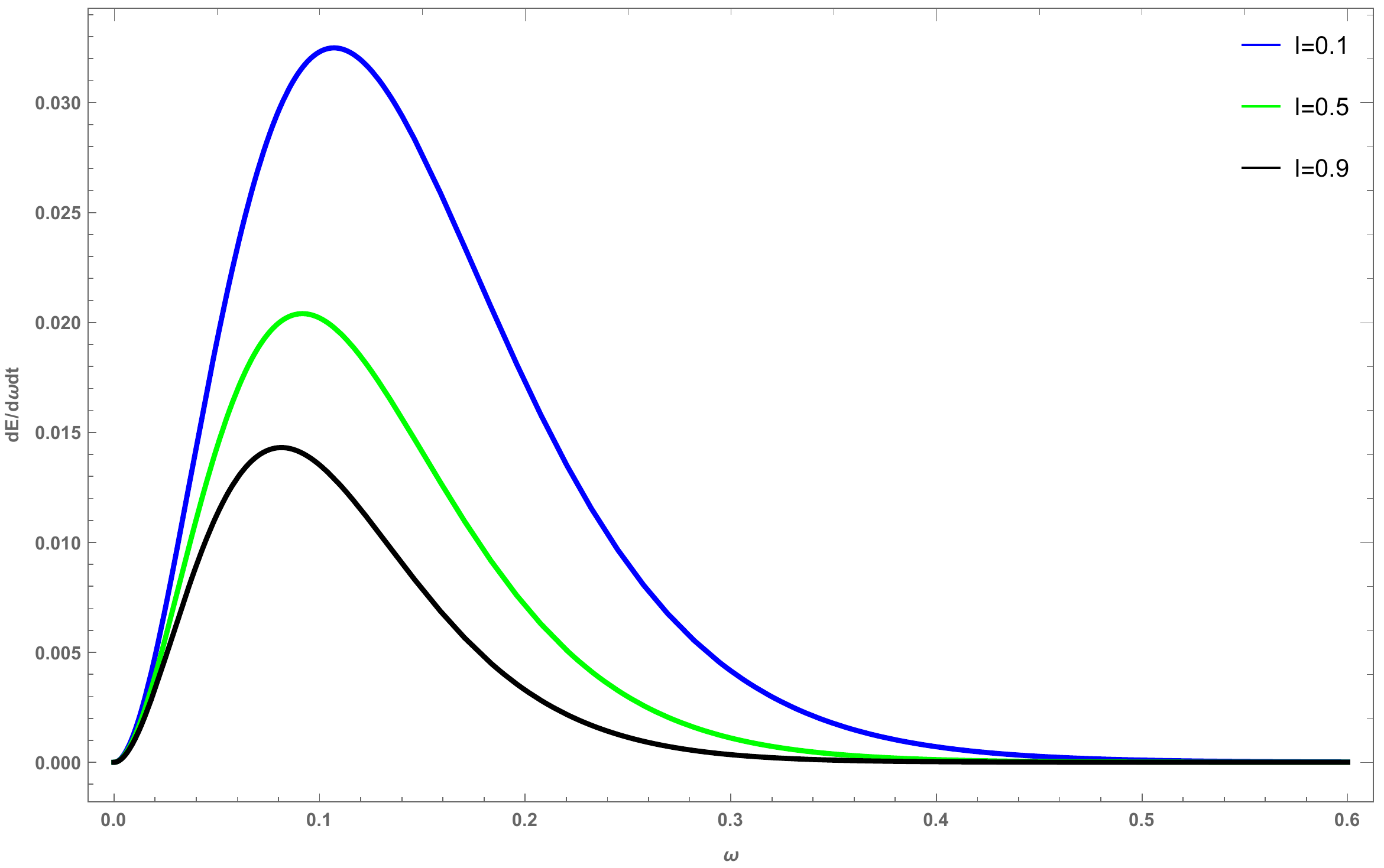} \caption{The BH emission rate for different values of LSB parameter $\ell$
at $M=1$ and $g=0.5$.}
\label{eml} 
\end{figure}

The plots in Fig. \ref{emg} and Fig. \ref{eml} shows the variation
of energy emission rate for frequency $\omega$ in different values
of the global monopole parameter $g$ and LSB parameter $\ell$, respectively.

\section{Weak deflection angle of Schwarzschild Like Solution with Global
Monopole in EHB Gravity}

To obtain the weak deflection angle of Schwarzschild Like Solution
with Global Monopole in Bumblebee Gravity, we write the metric in
the optical form within equatorial plane $\theta=\pi/2$, and obtain
null geodesics ($\mathrm{d}s^{2}=0$): 
\begin{eqnarray}
\mathrm{d}t^{2}=\frac{(1+\ell)\mathrm{d}r^{2}}{f(r)^{2}}+\frac{g^{2}r^{2}\mathrm{d}\phi^{2}}{f(r)},\label{metric2}
\end{eqnarray}
where $f(r)=1-\frac{2M}{r}$.

\begin{figure}[h!]
\center \includegraphics[width=0.5\textwidth]{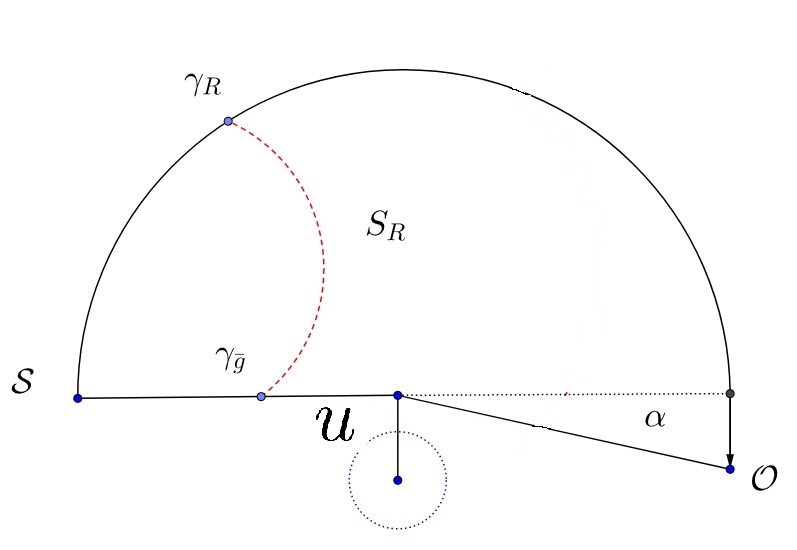} \caption{ Here, the integration domain $S_{R}$, namely the equatorial
plane $(r,\varphi)$, in which $\alpha$ is the total deflection angle
and $u$ is the impact parameter.}
\label{f1} 
\end{figure}

Afterwards, we calculate the Gaussian optical curvature $K$ in 2-dimensions
for the above space, which gives an intrinsic property of the space:
\begin{equation}
K=\frac{R_{icciScalar}}{2}\approx-2\,{\frac{M}{{r}^{3}}}+2\,{\frac{Ml}{{r}^{3}}}-2\,{\frac{M{l}^{2}}{{r}^{3}}}.\label{curvature}
\end{equation}
Note, that the Gaussian optical curvature is found as negative in
leading order terms, which imply that all the light rays locally diverge.
Hence, after converging, to obtain the multiple images, we will use
the Gauss-Bonnet theorem for the region  $S_{R}$ in manifold
shown in Fig. \ref{f1}, with boundary of circular curve $\gamma_{R}$
and geodesic curve $\gamma_{\bar{g}}$ as

$\partial S_{R}=\gamma_{\bar{g}}\cup\gamma_{R}$ \cite{Gibbons:2008rj}
\begin{equation}
\iint\limits _{S_{R}}K\,\mathrm{d}S+\oint\limits _{\partial S_{R}}\bar{\kappa}\,\mathrm{d}t=2\pi\chi(S_{R})-(\theta_{O}+\theta_{S})=\pi.\label{gaussbonnet}
\end{equation}

where the $\bar{\kappa}$ stands for the geodesics curvature. Moreover,
$\theta_{O}+\theta_{S}\to\pi$ implies that jump angles become $\pi/2$
when $R$ going to infinity. Also one can say that $S_{R}$ is non-singular
region, so that the Euler characteristic is $\chi(S_{R})=1$. Then,
$\bar{\kappa}(\gamma_{\tilde{g}})=0$. In addition, the near asymptotic
limit of $R$, $\gamma_{R}:=r(\varphi)=R=const.$, one can write the
radial component of the geodesic curvature as follows: \cite{Gibbons:2008rj}

\begin{equation}
\bar{\kappa}(\gamma_{R})=\left|\nabla_{\dot{\gamma}_{R}}\dot{\gamma}_{R}\right|=\left(\tilde{g}_{rr}\dot{\gamma}_{R}^{r}\dot{\gamma}_{R}^{r}\right)^{\frac{1}{2}}\to\frac{1}{R}.
\end{equation}

After substitute this result, yielding: $\bar{\kappa}(\gamma_{R})\mathrm{d}t=\frac{g}{\sqrt{(1+\ell)}}\,\mathrm{d}\,\varphi$.
Using the straight line approximation for the weak field regions,
$r_{s}=u/\sin\varphi$ at zeroth-order, where $u$ is the impact parameter.
Then the Gauss-Bonnet equation becomes: 
\begin{eqnarray}
\pi=\int\limits _{0}^{\pi}\int\limits _{\frac{u}{\sin\varphi}}^{\infty}K\,\mathrm{d}S+\frac{g}{\sqrt{(1+\ell)}}\int\limits _{0}^{\pi+\alpha}\mathrm{d}\varphi,\label{def1}
\end{eqnarray}
where optical surface area is defined as $\mathrm{d}S=r\mathrm{d}r\,\mathrm{d}\varphi$,
and $\alpha$ is a deflection angle.

Afterwards, the optical geometry \eqref{curvature} within the Gauss-Bonnet
theorem \eqref{def1}, give us the weak deflection angle: 
\begin{equation}
\alpha=\frac{\sqrt{(1+\ell)}\pi}{g}-\pi-\sqrt{(1+\ell)}\int\limits _{0}^{\pi}\int\limits _{\frac{u}{\sin\varphi}}^{\infty}K\,r\,\mathrm{d}r\,\mathrm{d}\varphi.\label{angle}
\end{equation}

Hence, the weak deflection angle $\alpha$ of Schwarzschild Like Solution
with Global Monopole in Bumblebee Gravity in weak field limits is
found as follows: 
\begin{equation}
\alpha\simeq4\,{\frac{M}{u}}+\,\frac{\ell\pi}{2}+2\,{\frac{M\ell}{u}}+\,\frac{\bar{\mu}\pi}{2}.\label{deflection}
\end{equation}

\begin{figure}[ht!]
\centering \includegraphics[scale=0.6]{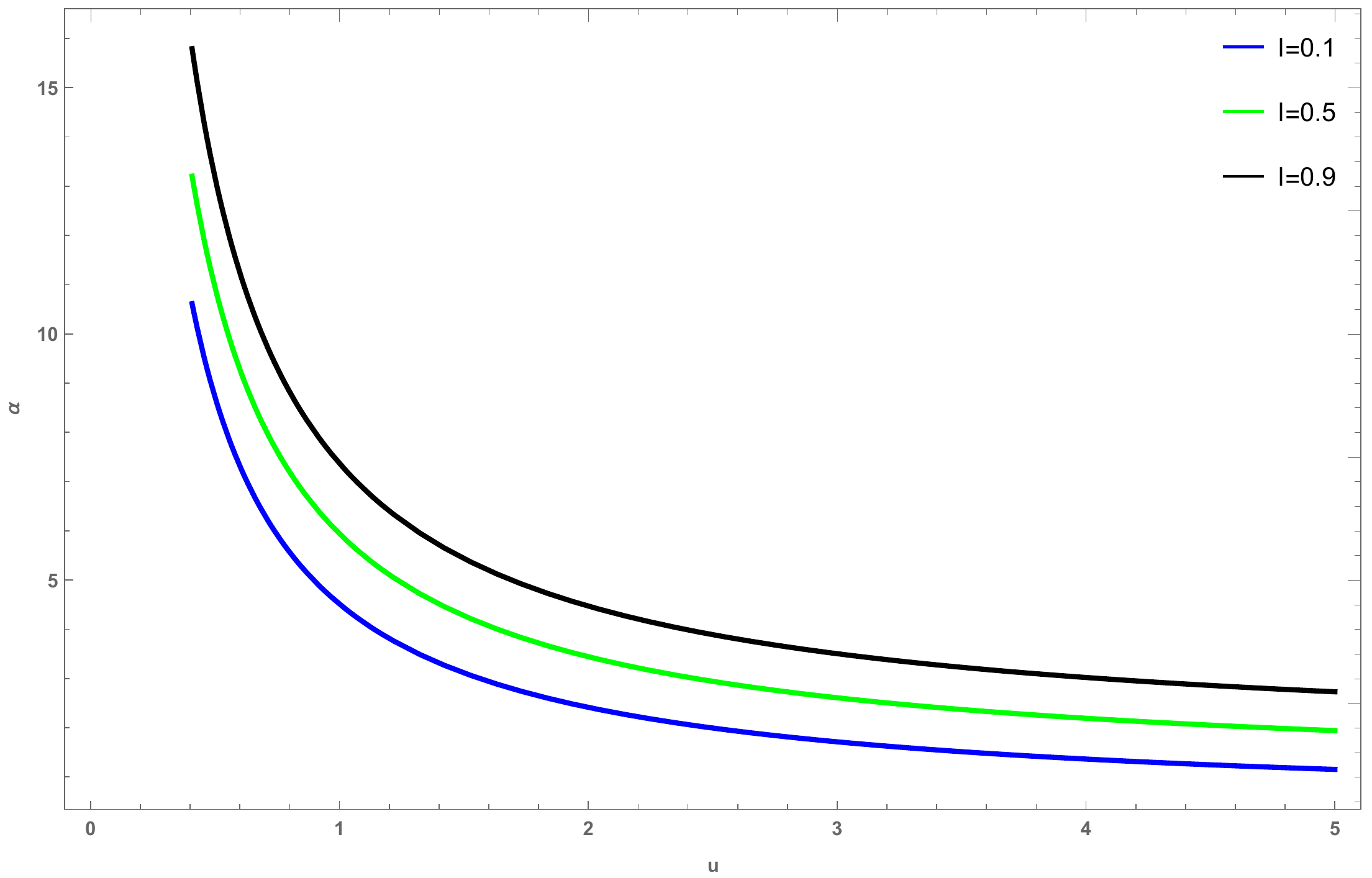} \caption{Weak deflection angle $\alpha$ versus impact parameter $u$ of Schwarzschild
Like Solution with Global Monopole in Bumblebee Gravity, for different
values of LSB parameter $\ell$ at $M=1$ and $\bar{\mu}=0.5$.}
\label{def2} 
\end{figure}

Note that the bumblebee parameter $\ell$, the mass term and the global
monopole term $\bar{\mu}$ all of them increase the deflection angle
as shown in Fig. \ref{def1} and Fig. \ref{def2}. The expression
of weak deflection angle is consistent with \cite{Li:2020dln} when
global monopole term $\bar{\mu}=0$. When the bumblebee parameter
$\ell=0$, the effect of global monopole fields on the deflection
angle in Eq. \ref{deflection} for comparison with other type of defects,
e.g., cosmic strings show that the increased value of the deflection
angle is too small, $\alpha\simeq\frac{4M}{b}+4\pi\nu$ \cite{Ovgun:2019wej}
with the order of 10 arcsec, (for cosmic string parameter $\nu\simeq10^{-6}$)
where the global monopole parameter is $\bar{\mu}\simeq10^{-5}$
\cite{Vilenkin:2000jqa}. Moreover, a  gravitating global monopole ($M \approx 0$)
deflects light although there is no gravity as $\alpha \simeq\,\,\frac{\ell\pi}{2}+\,\frac{\bar{\mu}\pi}{2} $.

\begin{figure}[ht!]
\centering \includegraphics[scale=0.6]{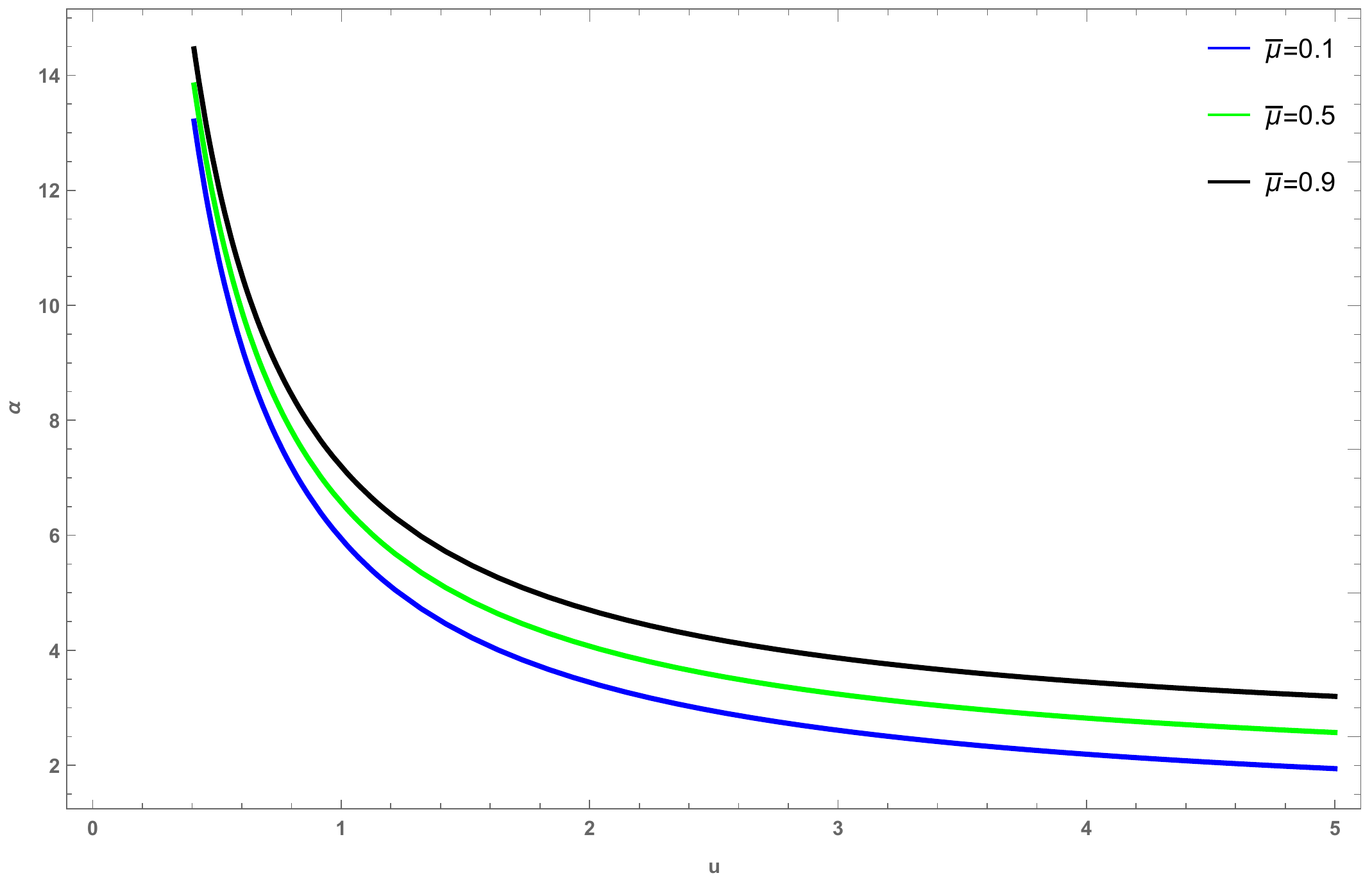} \caption{Weak deflection angle $\alpha$ versus impact parameter $u$ of Schwarzschild
Like Solution with Global Monopole in Bumblebee Gravity, for different
values of monopole parameter $\bar{\mu}$ at $M=1$ and $\ell=0.5$.}
\label{def1} 
\end{figure}

\section{Conclusions}

We have investigated a static spherically symmetric vacuum solution
for the EHB gravity in the presence of a global monopole. We have
obtained a spherically symmetric solution similar to the Schwarzschild
black hole. This solution reduces to spherically symmetric solution
in the LSB scenario when $\bar{\mu}=0$. Moreover, we have shown the
effect of the global monopole term on the shadow of the black hole
as well as the energy emission rate. The shadow radius of the black
hole increasing with the increasing value of global monopole parameter
$g$ as shown in Fig. \ref{shadow}. On the other hand, the shadow
radius does not depend on the LSB parameter $\ell$. Furthermore,
we have also calculated the deflection angle of the light in this
geometry by using the GBT. It is found that both the bumblebee parameter
and the global monopole term increases the deflection angle. If the
global monopole term is set to $\bar{\mu}=-\ell$ then part of the
effect of the bumblebee parameter on the deflection angle cancel out in leading order terms.
However, the bumblebee parameter still gives a correction to the deflection
angle. In addition to this, setting $\bar{\mu}=-\ell(1+{\frac{4M}{u\pi}})$
directly remove the effect of the bumblebee parameter on the deflection
angle. 
Note that our calculations was done in the weak-field limit, one can study the strong gravitational lensing to see other contributions.

In summary, we see that the global monopole term with the LSB parameter can be  altered the shadow of the black hole as well as the weak deflection angle.

The effect of the topological defects within the bumblebee gravity on the shadow radius has obtained for the first time in the literature here. We have showed that Schwarzschild-like black hole with a topological defect in bumblebee gravity's shadow can appear up to $5.5\%$ smaller than it would in a  case of Schwarzschild-like black hole, providing a clear way to test topological defect. While the current shadow image of black hole $M87^*$ is too fuzzy to tell the difference, efforts are underway to take even better pictures of more black holes, probing some of the deepest mysteries of the universe in the process. 

We end with some comments on future studies from several perspectives, which we did not address in this paper. We can extend the analysis for the generalized ansatz $b_{t}$ in Eq. \ref{b_m} can be generalized for  example $b_{t}=\text{constant}$. Second, One can also introduce $B_t=q=constant$  or $B^{\mu} B_{\mu}\neq0$ and try to find new black hole solutions. On the other hand, one can generalize EHB gravity to vector-tensor theories and so it will be of interest to study what happens in the presence of vector-tensor fields within the LSB parameter. We will leave these interesting things for future works.

\begin{acknowledgments}	
The authors are grateful to the anonymous referees for their valuable comments and suggestions to
improve the paper.
\end{acknowledgments}

\end{document}